\documentclass[prb,twocolumn]{revtex4-1} 

\usepackage{amsmath}  
\usepackage{amsfonts} 
\usepackage{graphicx} 
\usepackage{hyperref} 
\usepackage{pifont} 

\providecommand{\lightness}{\ensuremath{\lambda}}
\providecommand{\deltav}{\ensuremath{\Delta v}}
\providecommand{\ibf}{\ensuremath{f}}
\providecommand{\rinit}{\ensuremath{r_i}}
\providecommand{\vinit}{\ensuremath{v_i}}
\providecommand{\vafter}{\ensuremath{u_i}}
\providecommand{\vsail}{\ensuremath{v_s}}
\providecommand{\vspiral}{\ensuremath{v_{\rm spiral}}}
\providecommand{\rperi}{\ensuremath{r_p}}
\providecommand{\vperi}{\ensuremath{v_p}}
\providecommand{\vinf}{\ensuremath{v_\infty}}
\providecommand{\ms}{\ensuremath{{\rm m/s}}}
\providecommand{\kms}{\ensuremath{{\rm km/s}}}
\providecommand{\degrees}{\ensuremath{^\circ}}

\providecommand{\vinitvec}{\ensuremath{{\mathbf v}_i}}
\providecommand{\deltavvec}{\ensuremath{\Delta {\mathbf v}}}

\begin{document}

\title{The Sun Diver: Combining solar sails with the Oberth effect}

\author{Coryn A.L.\ Bailer-Jones}
\affiliation{Max Planck Institute for Astronomy, K\"onigstuhl 17, 69117 Heidelberg, Germany\\ \newline
{\rm Accepted to the} American Journal of Physics\\
{\rm 26 September 2020}\\}


\begin{abstract}
  A highly reflective sail provides a way to propel a spacecraft out of the solar system using solar radiation pressure.
  The closer the spacecraft is to the Sun when it starts its outward journey,
  the larger the radiation pressure and so the larger the final velocity.  For a spacecraft starting on the Earth's orbit, closer proximity can be achieved via a retrograde impulse from a rocket engine. The sail is then deployed at the closest approach to the Sun.  Employing the so-called Oberth effect, a second, prograde, impulse at closest approach will raise the final velocity further. Here I investigate
  how a fixed total impulse (\deltav) can best be distributed in this procedure to maximize the sail's velocity at infinity. Once \deltav\ exceeds a threshold that depends on 
  the lightness number of the sail (a measure of its sun-induced acceleration), the best strategy is to use all of the \deltav\ in the retrograde impulse to dive as close as possible to the Sun. 
 Below the threshold the best strategy is to use all of the \deltav\ in the prograde impulse and thus not to dive at all.
 Although larger velocities can be achieved with multi-stage impulsive transfers, this study shows some interesting and perhaps counter-intuitive consequences of combining impulses with solar sails.
  
\end{abstract}

\maketitle 

\section{Introduction}

Sailing on the Earth uses the pressure of the wind to propel a vehicle such as a ship.  The pressure come from material particles, namely air molecules, imparting momentum on the sail. Photons also possess momentum and therefore exert pressure, and these too can be used to propel a vehicle.  The pressure is much smaller, however, and it can only be used as an effective means of propulsion in the near-vacuum of space.

A photon of momentum $p$ has energy $E=pc$, where $c$ is the speed of light. As force equals rate of change of momentum, the force exerted by a photon hitting a surface is $\dot{E}/c$.
If a beam of photons of intensity $I$ is incident on a surface of area $A$, then $\dot{E}=IA$ so the pressure on that surface is $I/c$. This assumes the photons are absorbed. If they are instead perfectly reflected the pressure is $2I/c$. This is a small number: The intensity of the Sun incident on the top of the Earth's atmosphere is about 1370\,W\,m$^{-2}$, so photons reflected from a surface normal to the Sun's direction generate a pressure of just 9.1\,$\mu$Pa. Yet if we build a low mass spacecraft with a large reflecting surface, the
resulting acceleration is non-negligible, and as it is continuous, large velocities can be achieved. For example, if this pressure acted on a 100\,m${^2}$ sail of 1\,kg mass, the sail's velocity would change by 80\,\ms\ after one day (neglecting gravity).
This is the principle of a solar sail.

Solar sails are attractive because they free the spacecraft from having to carry propellant. It is an unavoidable consequence of the rocket equation\cite{walter2018, pinheiro2004} that the amount of propellant a rocket must carry to change its velocity by \deltav\ increases exponentially with \deltav. The reason is that most of the propellant is used to accelerate the unused propellant.

Solar sails are being considered as a way to explore the solar system.\cite{mcinnes1999,vulpetti2015}
A few prototypes have in fact been built and launched, after having been brought above the Earth's atmosphere by a conventional rocket.
Of particular interest is the Japanese mission IKAROS which used a 200\,m$^2$ sail of 16\,kg mass. It was launched in 2010 and flew past Venus, and is so far the only solar sail to have left Earth's orbit.\cite{tsuda2013}
Solar sails are particularly interesting for missions that require a large \deltav\ 
over a long duration, or that involve many maneuvers.
As sails provide continuous thrust, and can be tilted so that the net force on them is no longer directed along the line to the Sun, they can produce non-Keplerian orbits, enabling trajectories that would be much more expensive to attain with impulsive rocket maneuvers. Solar sails have also been investigated as a source of thrust for deep space and interstellar missions.\cite{liewer2000,lyngvi2005} Although the radiation pressure from the Sun drops with the inverse square of the distance from the Sun, so does its gravitational pull, so that if a solar sail is light enough it can escape the Sun's potential without any additional assistance.

It is this final application of solar sails -- attaining the largest possible velocity at infinity -- that we will investigate here. A large asymptotic velocity is paramount if we want to travel to the outer solar system or interstellar space in the shortest time possible.
We start from the realization that if a sail began its outward journey closer to the Sun than the Earth, it would gain extra acceleration due to the higher solar intensity in the first part of its journey, and so would achieve a larger velocity at infinity.\cite{mcinnes1999,gong2011} But a spacecraft starting from a circular orbit of 1\,au radius (1 astronomical unit, the mean Earth--Sun distance) would require an impulse in order to approach the Sun.
One way to achieve this is to use a rocket to apply a retrograde boost to decelerate the spacecraft by \deltav. This will put the spacecraft on an elliptical orbit that ``dives'' closer to the Sun (the larger \deltav, the closer the approach). Once the spacecraft reaches perihelion
it opens its sail and uses radiation pressure to sail away from the Sun.
But assuming that we have a fixed budget of \deltav\ available to change the velocity of the spacecraft, is the best strategy to use all of this to dive as close to the Sun as possible? The so-called Oberth effect,
explained in section~\ref{sec:oberth}, shows that applying the \deltav\ when the spacecraft is moving faster transfers more kinetic energy to the spacecraft than when it is moving slower.\cite{mcnutt2003} This suggests that it might be better to save some of the available \deltav\ for a prograde boost at perihelion, which is when the spacecraft is moving fastest.

We explore this scenario as a means of providing insight into the mechanics of solar sails and the use of impulsive boosts. The goal is not to identify the optimal set of orbital transfers that achieve the largest asymptotic velocity for a spacecraft. Such problems have been addressed in other articles and books.\cite{mcinnes1999,gong2011,vulpetti2015} Indeed, because sails can provide a continuous, variable, and non-central thrust, their orbits can be very complex, so we generally need sophisticated procedures and numerical methods to find the optimal trajectory for a given purpose.
This article provides instead an introduction to the topic by focusing on single-transfer orbits and analytic solutions, a topic that has not been covered comprehensively by other works.
Some of the results are counter-intuitive, thereby providing insight into both Keplerian orbits and solar sails.
Broad introductions to solar sailing are provided by Vulpetti et al.\cite{vulpetti2015} and, at a deeper level, McInnes.\cite{mcinnes1999}

We will make some simplifying assumptions. We assume the sail is a perfect reflector, and that the Sun is a point source that radiates isotropically.
We will also neglect the gravity of any body other than the Sun.  In practice a spacecraft launched from the Earth would need to escape the Earth's gravity, yet there are an infinite number of solar orbits it could be placed on in that process, each of which would require additional impulses.  Considering these would be important in practice, but here would only obfuscate the main issues.

It is well known from orbital mechanics that boosts tangential to the orbit are more efficient at changing the energy of the orbit than are boosts with a radial component.\cite{curtis2014,walter2018} For this reason we will only consider prograde boosts -- ones that increase the orbital velocity -- and retrograde boosts -- ones that decrease the orbital velocity. These boosts are assumed to be instantaneous.
All distances and velocities are relative to the Sun, except for \deltav, which is the change in velocity relative to the spacecraft's instantaneous reference frame. Velocities are non-relativistic, so a classical treatment suffices.

We first go over some background physics in section~\ref{sec:background_physics}, before exploring the nominal sun diver scenario in section~\ref{sec:nominal_scenario}. Some variations on this are considered in section~\ref{sec:other_scenarios}.

\section{Background physics}\label{sec:background_physics}

\subsection{Solar sails}\label{sec:solar_sails}

We consider a sail with its normal kept pointed at the Sun.
As noted in the previous section, the pressure of the solar photons is $2I/c$,
so the acceleration of a flat solar sail of mass per unit area $\sigma$ is $a = 2I/c\sigma$.
The solar intensity $I$ drops off with the inverse square of the distance $r$ from the Sun, so may be written
$I = L_s / 4 \pi r^2$, where $L_s$ is the luminosity of the Sun ($3.8 \times 10^{26}$\,W). We can then write $a = L_s / 2 \pi c \sigma r^2$. We will see momentarily that it is convenient to express this acceleration as a fraction of the local acceleration due to the Sun's gravity, $g=\mu/r^2$, where $\mu = GM_s$, $M_s$ is the mass of the Sun, and $G$ is the gravitational constant. For orientation, $g=5.9 \times 10^{-3}$\,m\,s$^{-2}$ when $r=1$\,au. The ratio $a/g$ is called the {\em lightness number} of the sail, \lightness, and it follows from the above that $\lightness = L_s / 2\pi c \mu \sigma$.
It is a property of the spacecraft and the Sun only, and in particular
is independent of $r$.
The acceleration of the sail may now be written $a = \lightness\mu/r^2$.

It is interesting to note in passing that we obtain $a=g$, i.e.\ $\lightness=1$, when the mass per unit area of the sail is $\sigma = L_s / 2\pi c \mu$. Numerically this is $1.5 \times 10^{-3}$\,kg\,m$^{-2}$, which is about ten times less than plastic food wrap. This indicates how small the solar radiation pressure is, and how light a solar sail needs to be to achieve an appreciable acceleration.

The sail experiences two forces: (i) the radiation pressure pushing it away from the Sun; (ii) gravity pulling it towards the Sun. Using $\boldsymbol{r}$ to denote the position vector of the sail relative to the Sun,
and $\hat{\boldsymbol{r}}$ to denote its unit vector, then from Newton's second law the
dynamical equation of the sail is
\begin{equation}
\ddot{\boldsymbol{r}} = -\frac{\mu}{r^2}\,\hat{\boldsymbol{r}} +  \frac{\lightness\mu}{r^2}\,\hat{\boldsymbol{r}} = -\frac{\mu(1-\lightness)}{r^2}\,\hat{\boldsymbol{r}} \ .
\end{equation}
This is the equation for Keplerian orbits in which the standard gravitational parameter is 
$\mu(1-\lightness)$ as opposed to $\mu$.
Thus for $0 < \lightness<1$ the effect of solar radiation pressure is equivalent to lowering the gravitational mass of the Sun. When $\lightness=1$ there is no net force, so the sail moves in a straight line or remains at rest. For $\lightness>1$ the net force is directed away from the Sun.

Consider a spacecraft initially on a circular orbit around the Sun. What orbit does the spacecraft acquire after it opens its sail and keeps its normal pointed at the Sun? The sail will now be moving faster than the circular orbital speed. 
For $0 < \lightness < 1/2$ it is straightforward to show that the spacecraft still has negative total energy, so its orbit becomes an ellipse.
For $\lightness = 1/2$ the total energy is zero, so the orbit is a parabola and it will reach infinity with zero velocity.
For $1/2 < \lightness < 1$ the total energy is positive, so the orbit is a hyperbola and the spacecraft is no longer bound to the Sun. For $\lightness>1$ the force is repulsive and the orbit is also a hyperbola, but now with the Sun at the other focus.

If a sail with $\lambda \geq 1/2$ is initially at a distance $\rinit$ from the Sun with a velocity $\vinit$, what velocity does it achieve at infinity (\vinf)? As the force is conservative the direction of \vinit\ is irrelevant, provided it is not directly towards the Sun.
Using conservation of energy we find
\begin{equation}
  \vinf^2 = \vinit^2 + \frac{2\mu(\lightness-1)}{\rinit} \ .
  \label{eqn:vinf_puresail}
\end{equation}
Clearly, the closer the sail is to the Sun initially, the larger $\vinf$ will be, provided $\vinit$ and/or $\lightness$ are large enough to permit escape at all. This suggests that to maximize $\vinf$ we should maneuver our spacecraft close to the Sun before opening its sail.


The lightness number \lightness\ is key to determining the velocity that the sail can achieve.
Sails launched to date had small lightness numbers.
The IKAROS\cite{tsuda2013} solar sail that went to Venus was primarily a technology demonstrator with a lightness number below 0.01.
LightSail 2, deployed into Earth's orbit in 2019 by a private organization (the Planetary Society),\cite{betts2019}
had a sail area of 32\,m$^2$ and mass of 5\,kg, giving it a theoretical lightness number of 0.01. One of the interstellar exploration concepts\cite{liewer2000} proposes a 400\,m diameter (125\,000\,m$^2$ area) sail weighing just 100\,kg. Together with a payload mass of 150\,kg, this implies a lightness number of 0.78.

\subsection{Oberth effect}\label{sec:oberth}

Consider a spacecraft travelling with a velocity \vinitvec\ in some inertial reference frame {\em S}.  The spacecraft's (specific) kinetic energy is $(1/2)|\vinitvec|^2$.  If it uses its rockets to increase its velocity by \deltavvec, its kinetic energy becomes $(1/2)|\vinitvec + \deltavvec|^2$. The increase in the spacecraft's kinetic energy is therefore $(1/2)|\deltavvec|^2 + \vinitvec \cdot \deltavvec$. This is maximized when \vinitvec\ and \deltavvec\ are parallel and then increases monotonically with increasing $|\vinitvec|$. That is, the faster the spacecraft is moving initially, the larger the increase in its kinetic energy for a given \deltav. Even though the spacecraft expends the same amount of energy {\em in its rest frame} to produce
a given \deltav, independent of \vinit, more of this energy goes into the kinetic energy of the spacecraft in {\em S} -- and thus less into the kinetic energy of the propellant -- the faster the spacecraft is moving in  {\em S}.

This observation can be exploited by a spacecraft to optimize the use of rocket propellant to escape from a gravitational field. If a spacecraft is on an elliptical orbit, then while its energy (kinetic plus potential) is conserved along the orbit, its velocity will vary, being largest at periapsis.\cite{periapsis_definition}
  Thus we will maximize the increase in the energy of the spacecraft if we apply the impulse \deltav\ at periapsis. Specifically, we fire the rockets tangentially to the orbit to increase its velocity -- a prograde boost.
This principle of maximizing kinetic energy increase is sometimes known as the Oberth effect, after the pioneering rocket scientist Hermann Oberth who first described it in the 1920s.\cite{oberth1929}
If a spacecraft has a velocity \vinit\ at a point on an elliptical orbit that is a distance \rinit\ from the central body, then if a prograde boost of \deltav\ is applied at periapsis where the velocity is \vperi, it is straightforward to show that the velocity $v_f$ of the spacecraft when it returns to \rinit\ (but now on a different orbit) is given by
\begin{equation}
  v_f^2 = \vinit^2 + (\deltav)^2\left(\frac{2\vperi}{\deltav} + 1\right) \ .
  \label{eqn:oberth}
\end{equation}
Clearly, the larger \vperi\ for given \vinit\ and \deltav, the larger $v_f$.

\section{Sun diver}\label{sec:nominal_scenario}

Equations~(\ref{eqn:vinf_puresail}) and (\ref{eqn:oberth}) suggest that we can maximize the velocity of our spacecraft at infinity if we maneuver our spacecraft as close to the Sun as possible before opening its sail and/or applying a prograde boost.  If our spacecraft starts on a circular orbit of radius $\rinit$ and velocity $\vinit$, we can use a retrograde boost to lower the orbital velocity and thus drop into an elliptical orbit with perihelion less than $\rinit$. This is the classic Hohmann transfer orbit.\cite{curtis2014} At perihelion we open the sail and apply an instantaneous prograde boost as motivated by the Oberth effect.

In practice our spacecraft will carry a fixed amount of propellant which, according to the rocket equation, corresponds to fixed total $\deltav$ budget. This presents us with a dilemma. Do we
\begin{itemize}
\item[(a)] Use the full $\deltav$ in the retrograde boost to drop as close as possible to the Sun, but leave no propellant for a prograde boost at perihelion, i.e.\ just rely on the sail from there?
\item[(b)] Forego the dive entirely and apply a full prograde boost on the initial circular orbit
  as we open the sail?
\end{itemize}
Or do we perform a combination of the two?
\begin{figure}[t]
\centering
\includegraphics[width=0.49\textwidth, angle=0]{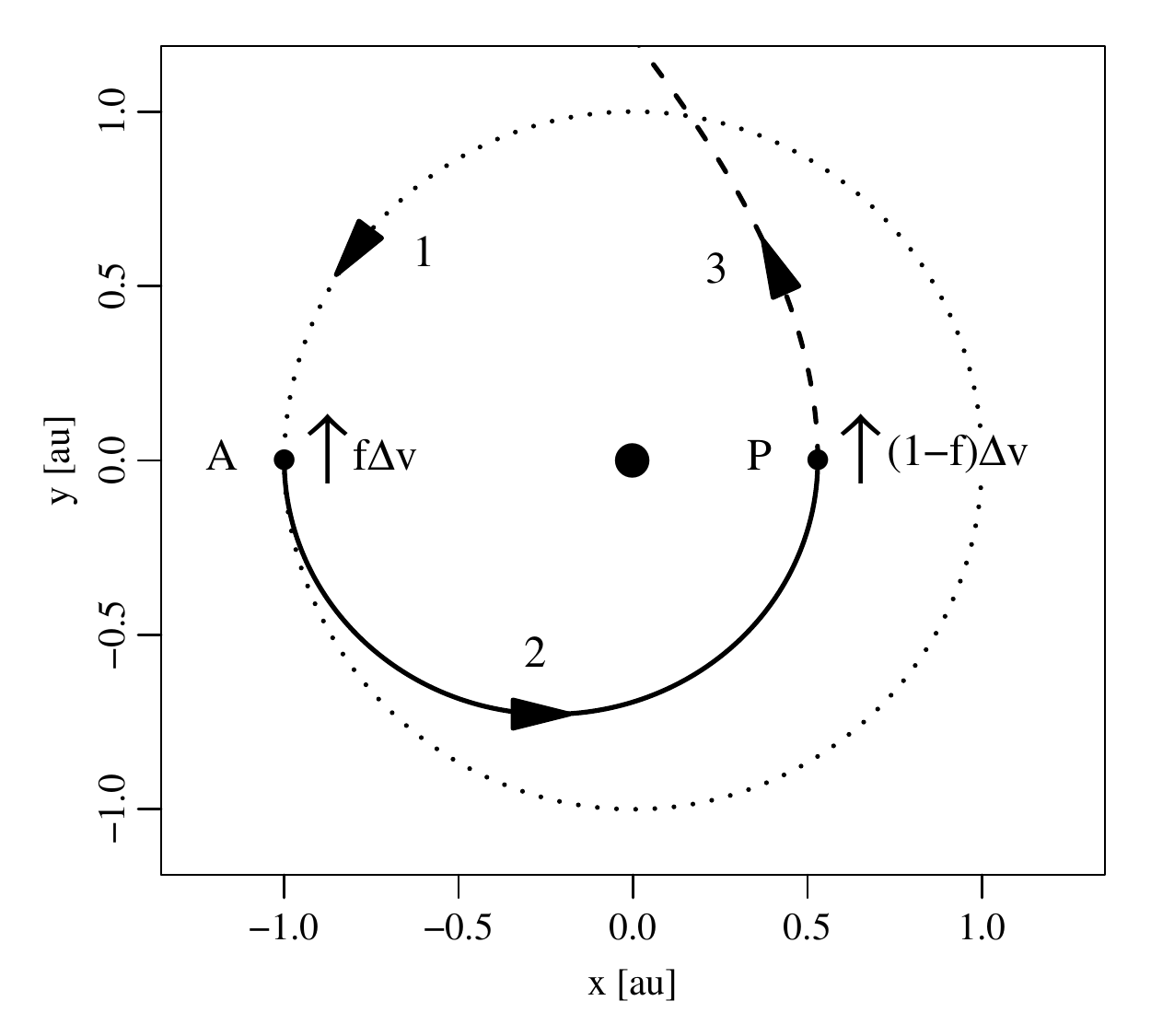}
\caption{The orbits in the nominal sun diver scenario: \ding{172} an initial circular orbit; \ding{173} an elliptical transfer orbit after the retrograde boost of $\ibf\deltav$ of A; \ding{174} the final orbit after the prograde boost of $(1-\ibf)\deltav$ and sail deployment at P. These orbits are shown to scale with $\rinit=1$\,au, $\deltav=10$\,\kms, $\ibf=0.5$, and $\lightness=0.3$, which makes orbit \ding{174} a hyperbola. The dot in the centre is the Sun (not to scale).}
\label{fig:nominal_scenario}
\end{figure}
This is the general case, shown in Fig.~\ref{fig:nominal_scenario}.
The spacecraft starts on the circular orbit \ding{172}.  With $0 \leq \ibf \leq 1$, a retrograde boost of $\ibf\deltav$ is applied at A to put the spacecraft on the elliptical transfer orbit \ding{173}. At this orbit's perihelion P, a prograde boost of $(1-\ibf)\deltav$ is applied and the solar sail is simultaneously deployed to put the spacecraft on orbit \ding{174}. 
It is not obvious what value of \ibf\ will produce the largest asymptotic velocity. The retrograde boost at A will {\rm reduce} the energy of the spacecraft, yet we hope to more than recover this from the higher intensity solar radiation closer to the Sun.



To resolve this we will now compute the velocity of the sail for the general case.
The initial circular orbit \ding{172} of radius \rinit\ has an orbital velocity \vinit\ given by
\begin{equation}
  \vinit^2 = \frac{\mu}{\rinit} \ .
  \label{eqn:vinit}
\end{equation}
At point A on this orbit the retrograde boost is applied, leaving the spacecraft with a velocity
\begin{equation}
  \vafter = \vinit - \ibf\deltav
  \label{eqn:vafter}
\end{equation}
which puts it on elliptical orbit \ding{173}.
Immediately after the boost the spacecraft has not yet moved, so its (specific) energy is
\begin{equation}
  E_{\rm A2} = -\frac{\mu}{\rinit} + \frac{1}{2}\vafter^2
  \label{eqn:E_A2}
\end{equation}
where the first subscript (A) refers to the position and the second subscript (2) to the orbit.
The spacecraft cruises from its aphelion at A to its perihelion at P, where the radius is \rperi, the velocity is \vperi, and the energy is
\begin{equation}
  E_{\rm P2} = -\frac{\mu}{\rperi} + \frac{1}{2}\vperi^2 \ .
  \label{eqn:E_P2}
\end{equation}
By conservation of energy $E_{\rm A2} = E_{\rm P2}$.
The (specific) angular momentum, 
${\boldsymbol r} \times {\boldsymbol v}$, is also conserved. Evaluating this at A and P
is simple because the velocity is perpendicular to the radial vector. Thus
\begin{equation}
  \rinit\vafter = \rperi\vperi \ .
  \label{eqn:L_conservation}
\end{equation}
Equating Eqs.~(\ref{eqn:E_A2}) and~(\ref{eqn:E_P2}) and substituting for \rperi\ from Eq.~(\ref{eqn:L_conservation}) gives a quadratic equation in \vperi,
\begin{equation}
  \vperi^2 - \frac{2\mu}{\rinit\vafter}\,\vperi + \frac{2\mu}{\rinit} - \vafter^2 = 0 \ .
  \label{eqn:vperi_quadratic}
\end{equation}
This has two solutions. One is $\vperi=\vafter$, which only makes sense when $\ibf=0$, i.e.\ the spacecraft makes no retrograde boost and so no dive. This can be treated as special case of the other solution, which is
\begin{equation}
  \vperi = \frac{2\mu}{\rinit\vafter} - \vafter \ .
  \label{eqn:vperi}
\end{equation}

When the spacecraft arrives at P on orbit \ding{173}, it simultaneously opens its sail and applies a prograde boost of $(1-\ibf)\deltav$, putting it onto orbit \ding{174}. As the sail is now open and its normal directed towards the Sun, the gravitational parameter is reduced by the factor $(1-\lightness)$ (section~\ref{sec:solar_sails}), so the spacecraft's energy is
\begin{equation}
  E_{\rm P3} = -\frac{\mu(1-\lightness)}{\rperi} + \frac{1}{2}[\vperi + (1-\ibf)\deltav]^2 \ .
  \label{eqn:E_P3}
\end{equation}
The spacecraft will cruise away from the Sun and will get at least as far as \rinit.
Its energy at some distance $r$ where its velocity is \vsail\ is 
\begin{equation}
  E_{\rm r3} = -\frac{\mu(1-\lightness)}{r} + \frac{1}{2}\vsail^2 \ .
  \label{eqn:E_r3}
\end{equation}
From conservation of energy we can equate $E_{\rm P3}$ and $E_{\rm r3}$ and then substitute for \vperi\ from Eq.~(\ref{eqn:vperi}). After a few lines of algebra we get the following expression for \vsail\ in terms of the initial orbit, \lightness, \deltav, and \ibf\ (via \vafter\ from Eq.~\ref{eqn:vafter}),
\begin{equation}
  \vsail^2 = 4\lightness\frac{\vinit^4}{\vafter^2} + 2\mu(1-\lightness)\left(\frac{1}{r} + \frac{1}{\rinit}\right)
  + (\vinit - \deltav)^2 - 4\frac{\vinit^2}{\vafter}(\vinit-\deltav) \ .
  \label{eqn:vsail}
\end{equation}
Note that \rinit\ and \vinit\ are not independent quantities: They are related
via Eq.~(\ref{eqn:vinit}) because we assumed a circular initial orbit when using angular momentum conservation.

Equation~(\ref{eqn:vsail}) is not very intuitive, but we can check that it gives the right results in certain limiting cases. For example, $\ibf=0$ corresponds to not doing any dive and applying the
full $\deltav$ in the initial orbit at the same time as the sail is deployed. Equation~(\ref{eqn:vsail}) then gives $\vsail^2 = (\vinit + \deltav)^2$
at $r=\rinit$ for all \lightness, as we would expect.

What value of \ibf\ gives the largest value of \vsail\ at some distance $r$? This is potentially a function of all the other parameters in Eq.~(\ref{eqn:vsail}).  We consider the spacecraft starting at the Earth's orbit of $\rinit=1$\,au, in which case $\vinit=29.8$\,\kms. While the numerical results will of course change when selecting a different initial orbit, the strategy that we should adopt to achieve the maximum velocity at infinity is independent of this.
Furthermore, it is sufficient to consider the velocity of the sail upon its return to $r=\rinit$, as we will see that the strategy which maximizes the velocity here will also maximize it as $r \rightarrow \infty$, if the spacecraft can reach infinity at all.

\begin{figure}[t]
\centering
\includegraphics[width=0.49\textwidth, angle=0]{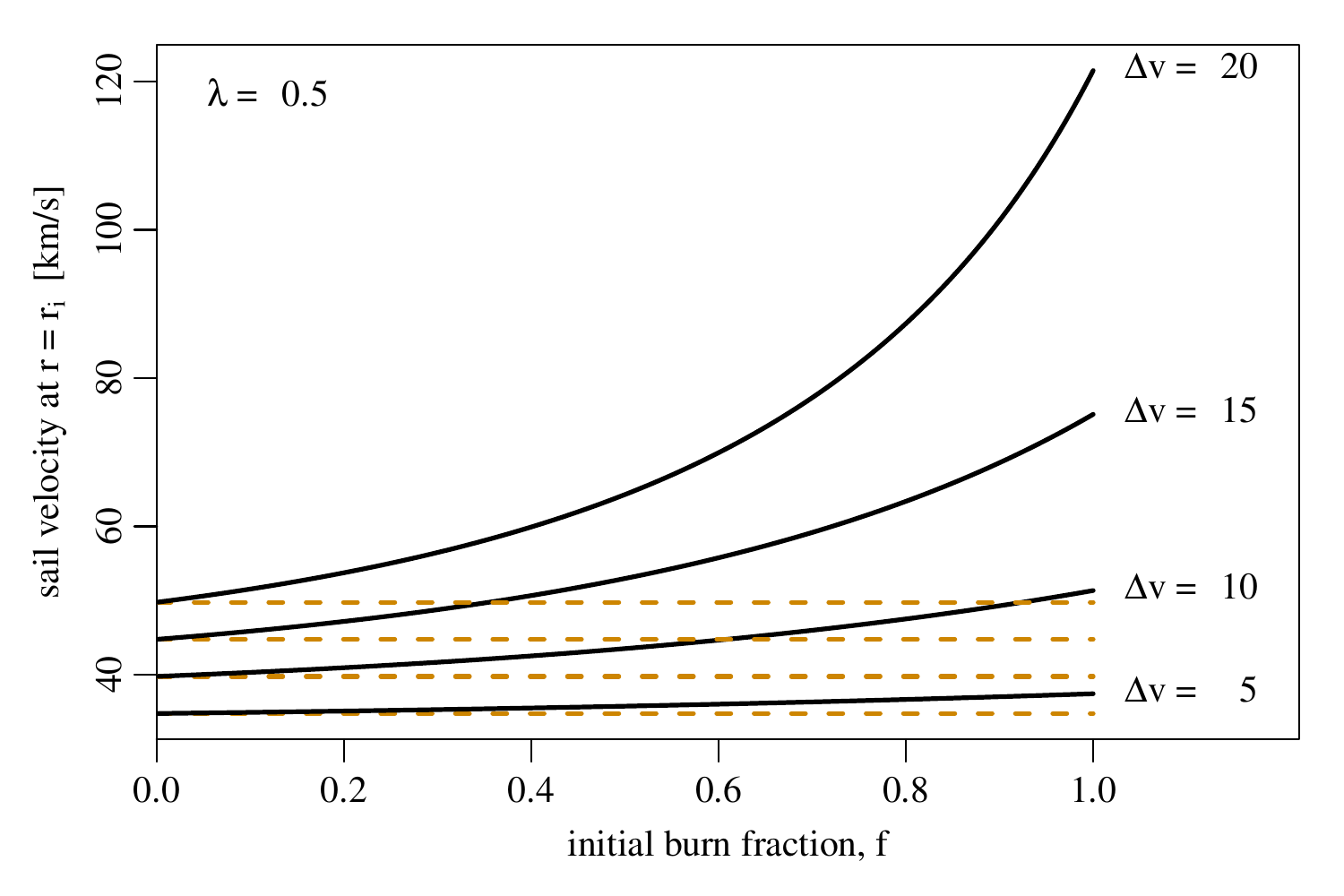}
\caption{The solid lines show the sail velocity \vsail\ from Eq.~(\ref{eqn:vsail}) for $r=\rinit$ (i.e.\ upon return to its initial altitude) and $\lightness=0.5$ as a function of \ibf\ for different values of \deltav\ (in \kms). The dashed lines show the corresponding sail velocity for $f=0$, i.e.\ when no dive is performed; these lines are horizontal.}
\label{fig:vrinit_vs_ibf_lightness0p5_various_delta}
\end{figure}
Figure~\ref{fig:vrinit_vs_ibf_lightness0p5_various_delta} shows how $\vsail(r\!=\!\rinit)$
varies with \ibf\ for a sail with $\lightness=0.5$ and various values\cite{deltav_magnitude} of \deltav. We see that for all $\deltav>0$, \vsail\ increases monotonically with \ibf.
In other words, the largest velocity is achieved by diving as close to the Sun as possible.
When $\deltav=0$ then of course $\vsail=\vinit$.

\begin{figure}[t]
\centering
\includegraphics[width=0.49\textwidth, angle=0]{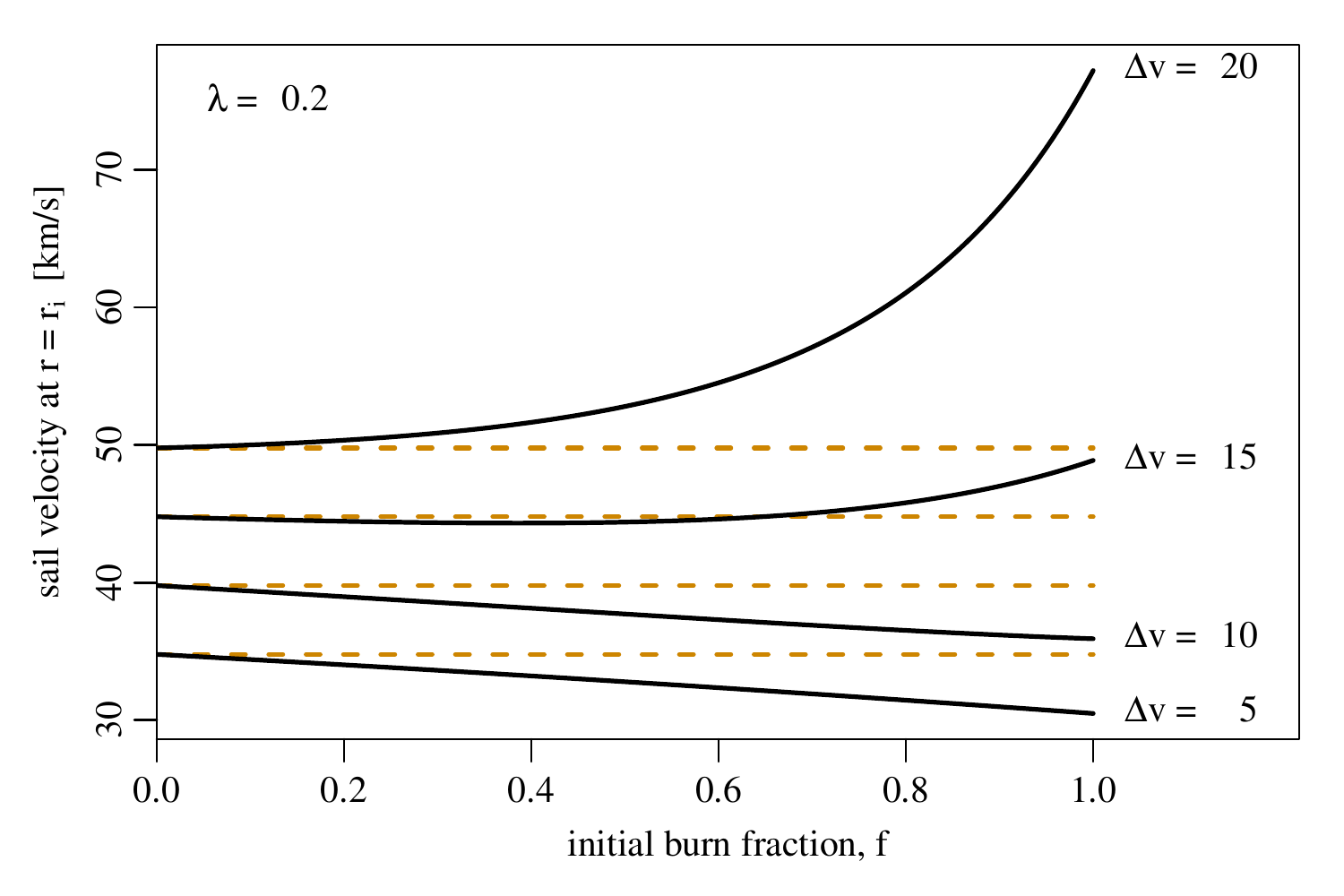}
\caption{As Fig.~\ref{fig:vrinit_vs_ibf_lightness0p5_various_delta} but for $\lightness=0.2$.}
\label{fig:vrinit_vs_ibf_lightness0p2_various_delta}
\end{figure}
Figure~\ref{fig:vrinit_vs_ibf_lightness0p2_various_delta} shows the same situation but for a smaller lightness number of $\lightness=0.2$. Now we see that for smaller values of $\deltav$, $\vsail(r\!=\!\rinit)$ {\em decreases} monotonically with increasing \ibf. In these cases, therefore, the largest velocity upon return to $r=\rinit$ is achieved by not diving at all and simply doing a prograde boost of \deltav\ from the initial orbit. This boost is formally applied at point P in Fig.~\ref{fig:nominal_scenario}, although of course it could be applied at any point on the initial orbit.
It appears that for a given \lightness, there is a value of \deltav\ below which the best strategy is to apply the full boost prograde ($\ibf=0$, no dive) and above which the best strategy is to apply the full boost retrograde ($\ibf=1$, full dive).

\begin{figure}[t]
\centering
\includegraphics[width=0.49\textwidth, angle=0]{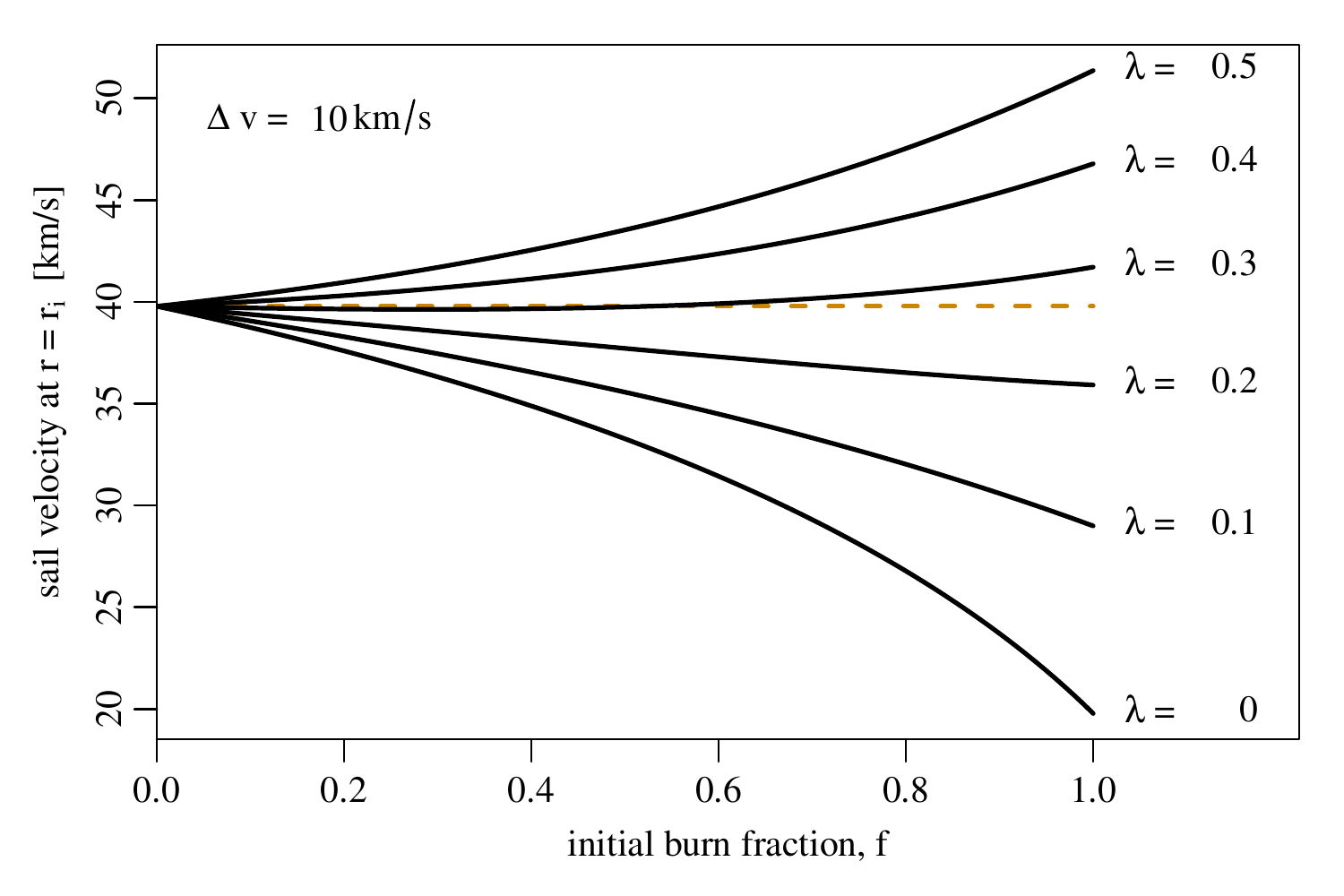}
\caption{As Fig.~\ref{fig:vrinit_vs_ibf_lightness0p5_various_delta} but now for $\deltav=10$\,\kms\ for lightness numbers ranging from 0.1 to 0.5. The dashed line is for $f=0$, i.e.\ no dive performed, which is the same for all lightness numbers.}
\label{fig:vrinit_vs_ibf_deltaV10_various_lightness}
\end{figure}
\begin{figure}[t]
\centering
\includegraphics[width=0.49\textwidth, angle=0]{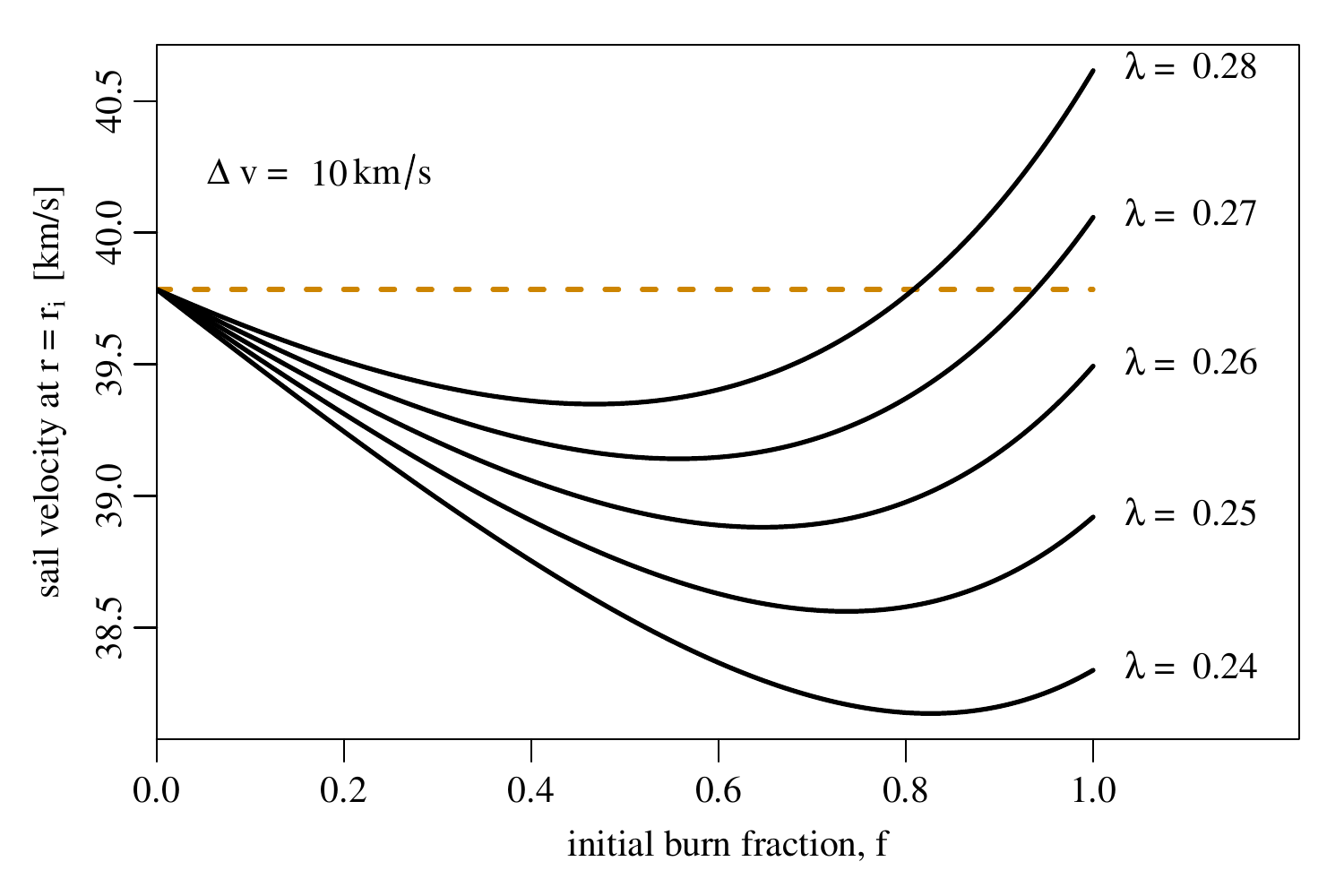}
\caption{As Fig.~\ref{fig:vrinit_vs_ibf_deltaV10_various_lightness} but for a narrower range of lightness numbers.}
\label{fig:vrinit_vs_ibf_deltaV10_various_lightness_2}
\end{figure}
This conclusion can also be obtained from Fig.~\ref{fig:vrinit_vs_ibf_deltaV10_various_lightness},
which plots $\vsail(r\!=\!\rinit)$ for $\deltav=10$\,\kms\ and lightness numbers from 0.0 to 0.5, and moreover in Fig.~\ref{fig:vrinit_vs_ibf_deltaV10_various_lightness_2} which zooms in on the transition region for lightness numbers between 0.24 and 0.28. The curves in the latter figure all show a minimum.
It can be shown by differentiation that if the function \vsail(\ibf)\ has a turning point at all within the range $0 \leq \ibf \leq 1$ (and it does not always have one, e.g.\ Fig.~\ref{fig:vrinit_vs_ibf_lightness0p5_various_delta}), then this is always a minimum. This is true for any $r$, and indeed any values of the other parameters.
Hence there is never an intermediate value of \ibf\ which will maximize \vsail: The optimum strategy is either $\ibf=0$ or $\ibf=1$. In other words, to achieve the maximum velocity at infinity we must use all the propellant in one go, either all at P ($\ibf=0$) or all at A ($\ibf=1$) in Fig.~\ref{fig:nominal_scenario}. Partitioning the propellant use between these points (or indeed any others points) will yield a lower velocity at infinity.
Which of the two strategies we should adopt depends on the values of the parameters.
For example, in Fig.~\ref{fig:vrinit_vs_ibf_deltaV10_various_lightness_2} we see that
for the three smallest values of \lightness\ shown (0.24, 0.25, 0.26), $\ibf=0$ maximizes \vsail, whereas for the two largest values of \lightness\ shown (0.27 and 0.28), $\ibf=1$ maximizes \vsail.

We can use an inequality to determine the relationship between the parameters that governs the transition from $\ibf=0$ to $\ibf=1$ being the best strategy. We ask for what values of the parameters is
\begin{equation}
\vsail^2(\ibf\!=\!1) > \vsail^2(\ibf\!=\!0) \ .
\end{equation}
Using Eq.~(\ref{eqn:vsail}) and a little manipulation this inequality can be written
\begin{equation}
  \lightness > \frac{(1-\deltav/\vinit)^2}{2-\deltav/\vinit} \ . 
  \label{eqn:transition}
\end{equation}
This holds for all $r$ because the term involving $r$ in Eq.~(\ref{eqn:vsail}) does not have a factor of \ibf\ in it, so cancels out.
\begin{figure}[t]
\centering
\includegraphics[width=0.49\textwidth, angle=0]{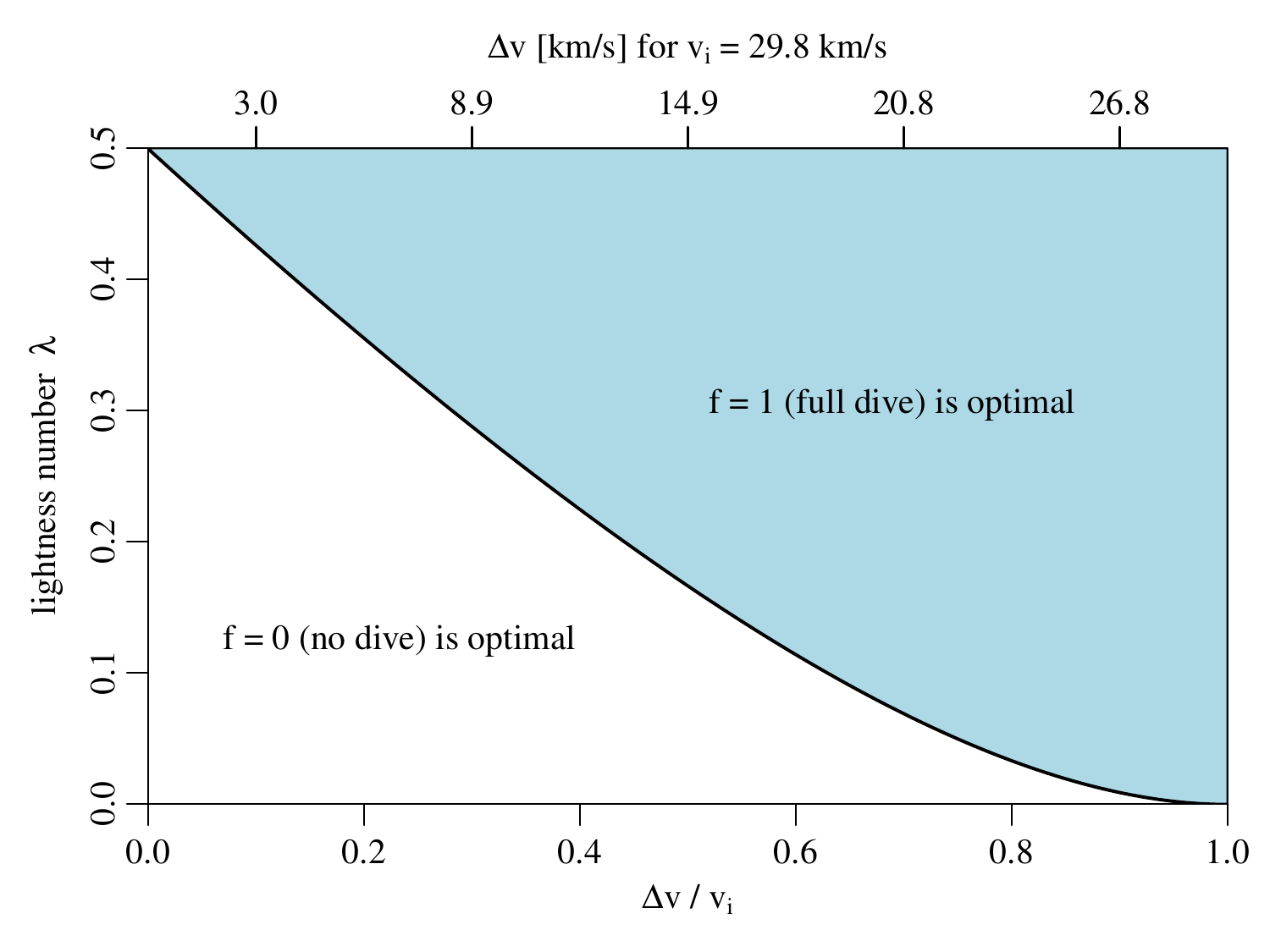}
\caption{The full retrograde boost ($\ibf=1$, full dive) achieves a larger sail velocity
\vsail\ than the full prograde boost ($\ibf=0$, no dive) for values
of \lightness\ and $\deltav/\vinit$ above the line (inequality~(\ref{eqn:transition})).
The opposite is true for values below the line.}
\label{fig:transition_lightness_vs_deltav}
\end{figure}
This inequality is plotted in Fig.~\ref{fig:transition_lightness_vs_deltav}, and is one of the main conclusions of this study. It shows that for a given \deltav\ the optimal strategy is not to dive if \lightness\ is too small.  The reason is that with a smaller lightness number, the extra kinetic energy provided by the solar radiation from moving close to the Sun does not compensate for the energy lost by performing the dive.

\begin{figure}[t]
\centering
\includegraphics[width=0.49\textwidth, angle=0]{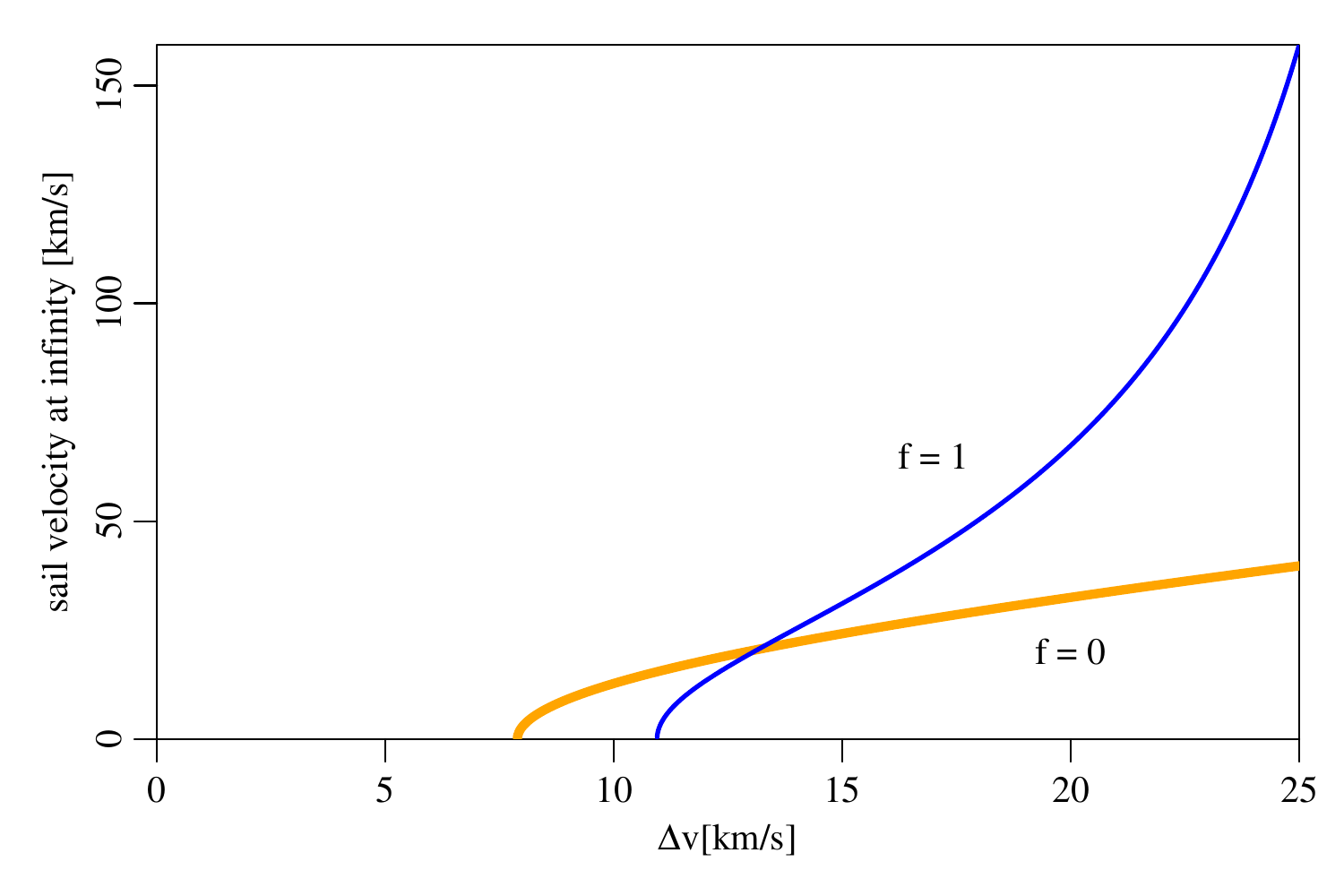}
\caption{The velocity of the spacecraft at infinity as a function of \deltav\ with $\lightness=0.2$ for
no dive ($\ibf=0$, thick line) and a full dive ($\ibf=1$, thin line).}
\label{fig:vrinf_vs_deltaV_ibf_0_1}
\end{figure}
Using Eq.~(\ref{eqn:vsail}) we can compute the velocity the sail will attain as $r \rightarrow \infty$. This is shown
for case $\lightness=0.2$ in Fig.~\ref{fig:vrinf_vs_deltaV_ibf_0_1} as a function of \deltav\ for
no dive and for a full dive. The transition between the strategies yielding the larger velocity occurs at $\deltav=13.2$\,\kms\ in accordance with inequality~(\ref{eqn:transition}). Note that if \deltav\ is too small the spacecraft cannot reach infinity at all.

\begin{figure}[t]
\centering
\includegraphics[width=0.49\textwidth, angle=0]{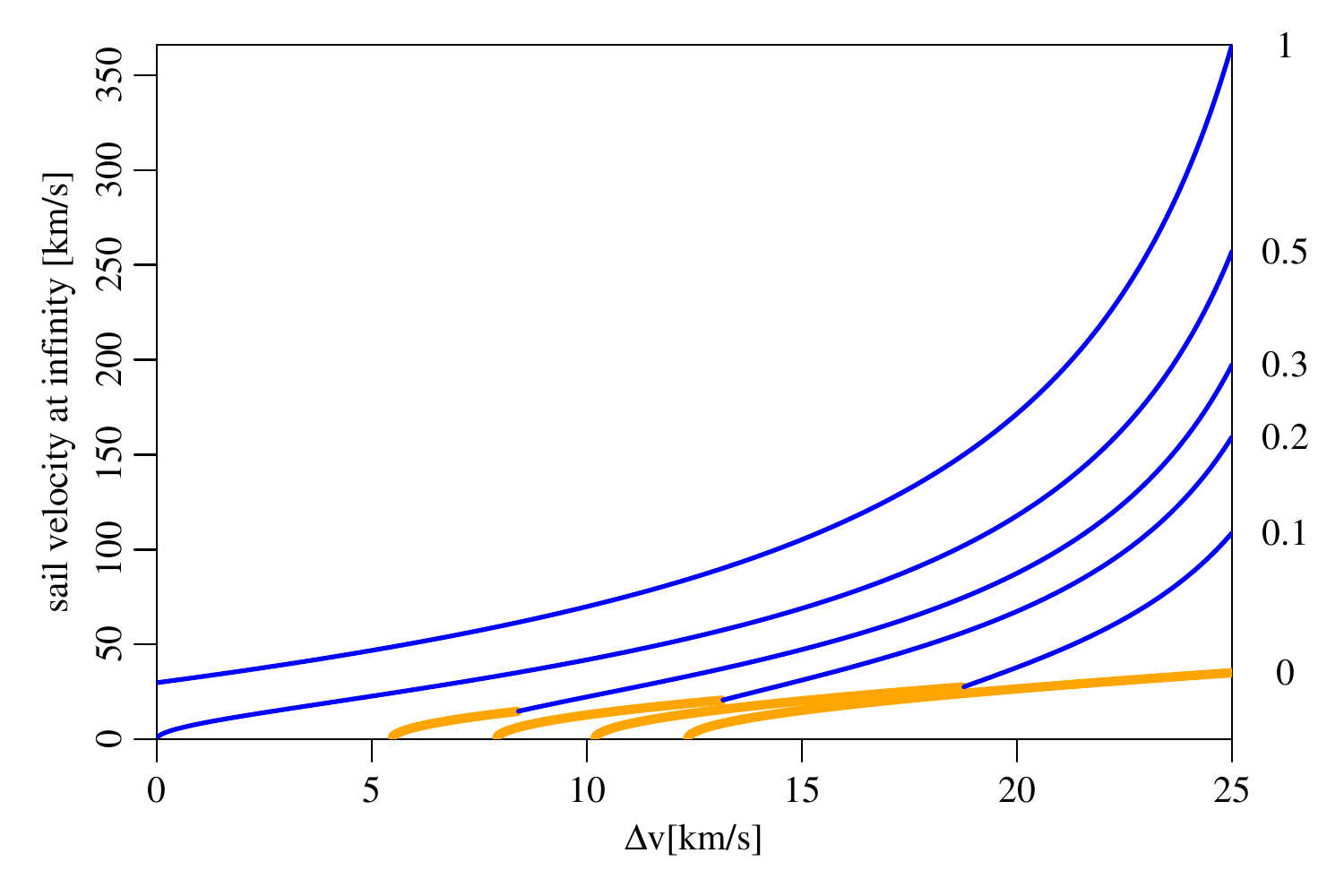}
\caption{The velocity of the spacecraft at infinity as a function of $\deltav$ for various lightness numbers \lightness\ (indicated on the right). For each combination of \deltav\ and \lightness\ the optimal boost strategy is selected according to Fig.~\ref{fig:transition_lightness_vs_deltav}.
  The line is thin where the optimal strategy is $\ibf=1$ (full dive) and thick where it is $\ibf=0$ (no dive).}
\label{fig:vrinf_vs_deltaV_various_lightness_optimal_ibf}
\end{figure}
Figure~\ref{fig:vrinf_vs_deltaV_various_lightness_optimal_ibf} shows the velocity at infinity for several different lightness numbers.
In each case the optimal strategy ($\ibf=0$ or $1$) at each \deltav\ has been adopted according to inequality~(\ref{eqn:transition}).
Only if $\lightness>0.5$ will the spacecraft reach infinity even for $\deltav=0$.
As we saw before, for $\lightness<0.5$ the optimal strategy for smaller values of \deltav\ is not to dive at all.
Recall that $\lightness=0$ corresponds to no sail.
%

\begin{figure}[t]
\centering
\includegraphics[width=0.49\textwidth, angle=0]{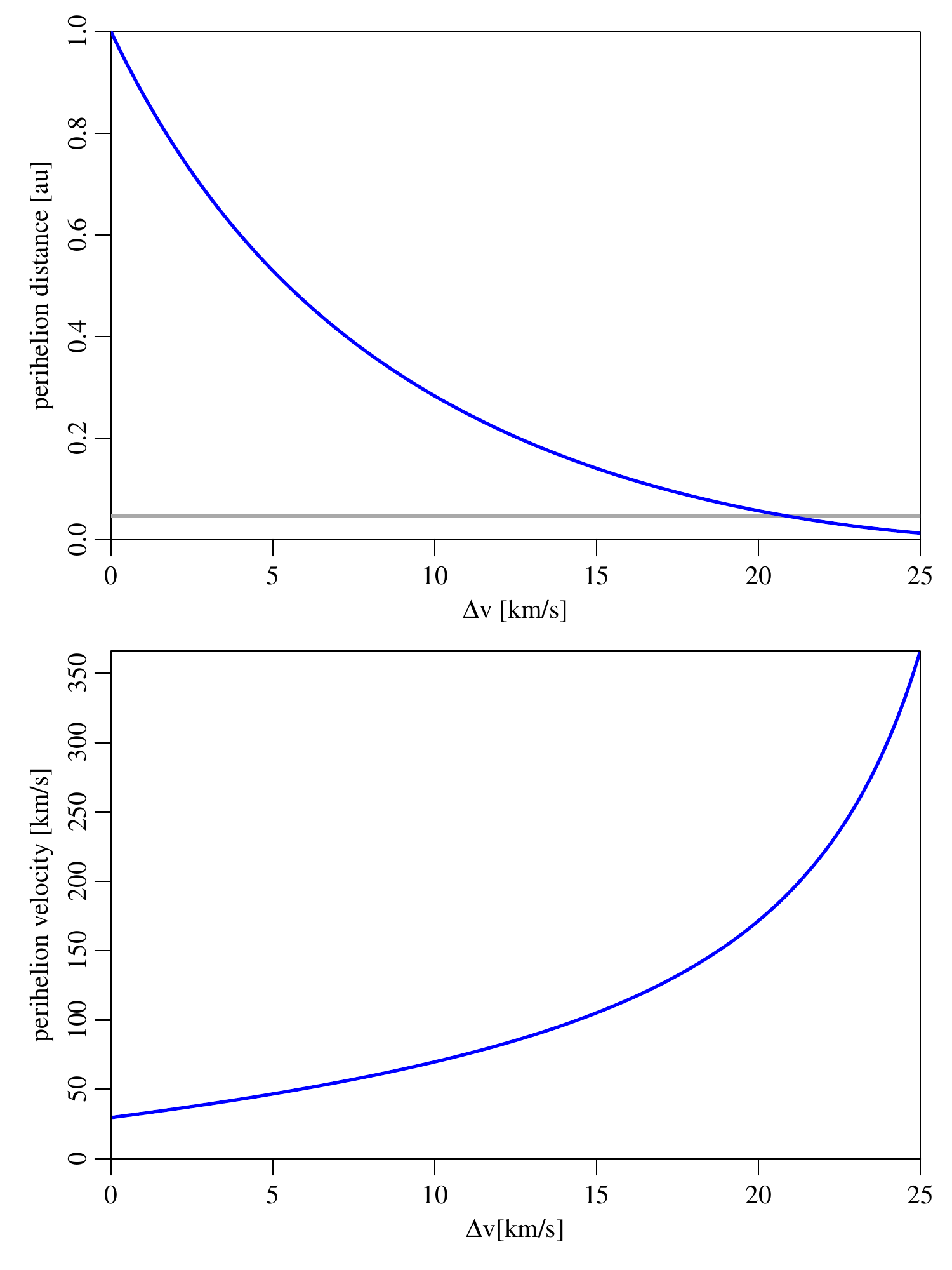}
\caption{Variation of the perihelion distance (top) and perihelion velocity (bottom) as a function of the size of the retrograde boost for a full dive ($\ibf=1$). The horizontal line in the upper panel indicates 10 solar radii.}
\label{fig:rperi_and_vperi2_vs_deltaV}
\end{figure}
How fast and close to the Sun does the spacecraft get in a full dive? This is shown in Fig.~\ref{fig:rperi_and_vperi2_vs_deltaV} as a function of \deltav.
For \deltav\ larger than about 20\,\kms\ the spacecraft will approach within 10 solar radii (0.047\,au), which is approximately the closest a spacecraft has ever approached the Sun (the Parker Solar Probe, although it achieved this through a series of gravity assists).\cite{longcope2000,psp}
Thermal considerations, i.e.\ not melting the spacecraft, would probably set the limit on the closest approach.

We have only considered $\deltav < \vinit$ in order to avoid the singularity caused by the spacecraft dropping into the centre of the Sun. If we had enough propellant for a larger boost, then the optimal strategy would be to dive as close to the Sun as possible and to use all of the remaining propellant in the prograde boost at perihelion. We will nonetheless look at the idea of applying prograde boosts higher in the potential in the next section.

\section{Variations on the sun diver scenario}\label{sec:other_scenarios}

\subsection{Boost at infinity}\label{sec:boost_at_infinity}

In the nominal scenario in the previous section we considered applying rocket boosts in only two places, namely on the initial circular orbit and/or at perihelion.  We chose perihelion because the Oberth effect tells us to apply the boost when the spacecraft is moving fastest. Yet when $\lightness>1$ the net force on the spacecraft is outwards, so the spacecraft is moving fastest when it reaches infinity. Is infinity therefore a more efficient place to apply $\deltav$ in this case? We saw that in order to achieve the largest final velocity when $\lightness>1/2$ we should do a full dive (Fig.~\ref{fig:transition_lightness_vs_deltav}).
So it is not immediately obvious which of the two following scenarios gives the largest velocity at infinity:
\begin{itemize}
\item[(a)] Apply the full $\deltav$ retrograde to do a full dive (same scenario (a) as in section~\ref{sec:nominal_scenario}).
\item[(c)] Open the sails on the initial circular orbit to sail to infinity, then apply the full $\deltav$.
\end{itemize}
The velocity at infinity for scenario (a) we obtain from Eq.~(\ref{eqn:vsail}) with $r \rightarrow \infty, \ibf=1$.
The velocity at infinity for scenario (c) before we apply the final boost is obtained from Eq.~(\ref{eqn:vsail}) with $r \rightarrow \infty, \deltav=0$.
We then add \deltav\ to get
\begin{equation}
  v_c = \vinit\sqrt{2\lightness-1} + \deltav \hspace*{1cm}\textrm{Scenario (c)} \ .
  \label{eqn:vsail_inf_c}
\end{equation}
\begin{figure}[t]
\centering
\includegraphics[width=0.49\textwidth, angle=0]{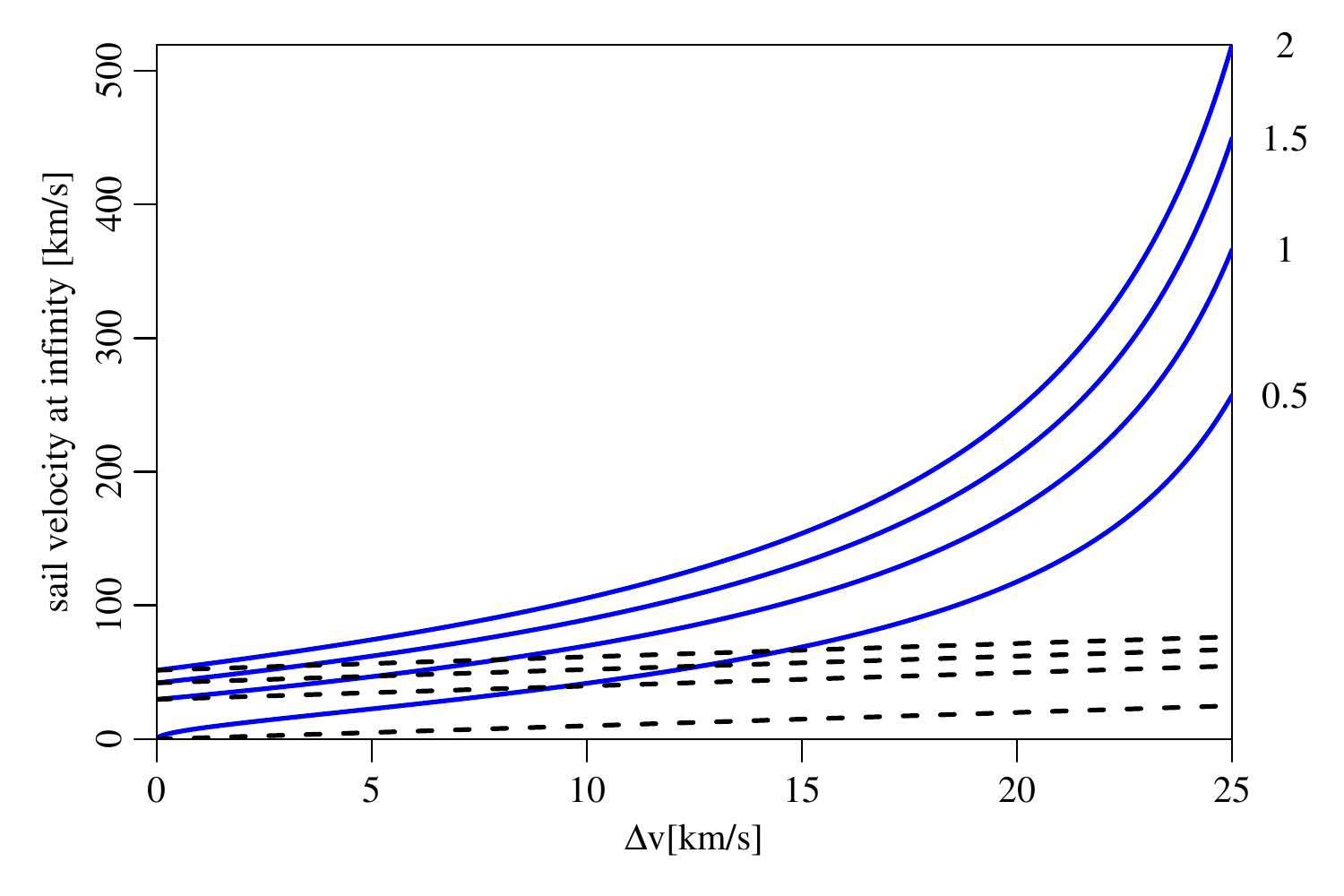}
\caption{The solid lines show the velocity of the spacecraft at infinity after performing a full dive (Eq.~(\ref{eqn:vsail}), $\ibf=1$) as a function of \deltav\ for various lightness numbers \lightness\ (indicated on the right). The dashed lines show the corresponding velocity achieved if no dive is performed, and the \deltav\ is applied at infinity instead (Eq.~(\ref{eqn:vsail_inf_c})).}
\label{fig:vrinf_vs_deltaV_various_lightness_ibf1_compare_boost_infinity}
\end{figure}
A comparison of the expressions for the two velocities is not very informative, but a plot makes it clear which scenario is superior. The two velocities are shown in Fig.~\ref{fig:vrinf_vs_deltaV_various_lightness_ibf1_compare_boost_infinity}
as a function of \deltav\ for various lightness numbers.
We only show $\lightness>1/2$ because for smaller lightness numbers the sail in scenario (c) cannot reach infinity.
It is clear from the plot that scenario (a) is superior to scenario (c), except at $\deltav=0$ where they are of course equivalent. The reason is that for such large lightness numbers, more energy is gained from the Sun by diving close to the Sun, than is lost by performing this dive.

In practice a significant part of the mass of a rocket is its propellant. Thus the mass of the spacecraft, and therefore its lightness number, depends strongly on whether the propellant has been expended.
In our nominal scenario, which includes scenario (a), the sail is only deployed after all the propellant has been depleted -- and the propellant tanks and engines would be jettisoned too -- so the spacecraft would have a small mass and thus large lightness number. Scenario (c), in contrast, requires the solar sail to accelerate all of the propellant to infinity, and so the spacecraft with the same sail would have a smaller lightness number than in (a). So in practice scenario (c) would be even worse.


\subsection{Sail towards the Sun}\label{sec:sail_to_sun}

If the normal of the sail is not kept parallel to the radial vector pointing from the Sun to the spacecraft, solar photons will exert a non-central force on the spacecraft. This leads to non-Keplerian orbits, the properties of which depend on how the pitch angle $\alpha$ between the radial and normal vectors varies. One particularly interesting solution to the dynamical equations, and one of the few analytic ones, occurs when $\alpha$ is kept fixed. This gives rise to {\em logarithmic spiral orbits}\cite{bacon1959} which are described in polar coordinates $(r,\theta)$ as
\begin{equation}
  r = \rinit\exp(\theta\tan\gamma)
  \label{eqn:rspiral}
\end{equation}
where $\gamma$ is a quantity (the spiral angle) that depends on $\alpha$ and \lightness\ only.\cite{mcinnes1999}
The sail describes a spiral path around the Sun.
The velocity of the sail, \vspiral, at a point $(r,\theta)$ in its orbit is
\begin{equation}
  \vspiral^2 = \frac{\mu}{r}[1 - \lightness\cos^2\!\alpha(\cos\alpha - \sin\alpha\tan\gamma)]
  \label{eqn:vspiral}
\end{equation}
where the radial and tangential components are $\vspiral\sin\gamma$ and $\vspiral\cos\gamma$ respectively.
This trajectory is interesting for our application because if $\alpha$ is negative, the solar photons act to decelerate the spacecraft compared to the non-sail Keplerian orbit, and so the spacecraft will spiral {\em inwards} towards the Sun. This is achieved without expending any propellant, and so allows us to apply the entire \deltav\ at perihelion. Can this be used to achieve a larger velocity at infinity than
the full dive scenario of section~\ref{sec:nominal_scenario}?
We compare the following two scenarios, both of which start from a circular orbit with the sail folded away.
\begin{itemize}
\item[(a)] Apply the full \deltav\ retrograde on the initial orbit in order to dive to a distance $r=\rperi$,
at which point we open the sail and then keep it pointed at the Sun (same scenario (a) as in section~\ref{sec:nominal_scenario}); 
\item[(d)] Tilt the sail in order to spiral towards the Sun until distance $r=\rperi$,
at which point we simultaneously apply the full \deltav\ prograde and turn the sail to keep it pointed at the Sun.
\end{itemize}

This is not an ideal comparison because the velocity on the spiral orbit in (d) immediately after the sail has been tilted is not equal in magnitude or direction to the velocity of the initial circular orbit. An additional impulse or maneuver would therefore be required to put the sail onto the spiral trajectory.
We can ignore this, however, because we will see that it does not change the answer to the above question.


To see which of these scenarios give us the largest velocity at infinity (or indeed any distance $r>\rperi$) we compare the velocities at $r=\rperi$, which are \vperi\ from Eq.~(\ref {eqn:vperi}) with $\vafter = \vinit -\deltav$ for scenario (a), and $\vspiral+\deltav$ from Eq.~(\ref{eqn:vspiral}) for scenario (d). Let us refer to these as the ``closest approach velocities''.
A comparison at $r=\rperi$ is sufficient because in both scenarios the force experienced by the spacecraft after closest approach is the same and is conservative (gravity plus photon pressure, both directed radially). The magnitude of the velocity at any later point is therefore determined entirely by the energy at $r=\rperi$.

By equating Eqs.~(\ref{eqn:E_A2}) and~(\ref{eqn:E_P2}) and eliminating \vafter\ using Eq.~(\ref{eqn:L_conservation}),
we may express \vperi\ in scenario (a) in terms of \rperi\ and \rinit\ only
\begin{equation}
  \vperi^2 = 2\mu\frac{\rinit}{\rperi}\frac{1}{(\rinit + \rperi)} \ .
  \label{eqn:vperi_alt}
\end{equation}
Although we are only interested in a full dive here, this expression actually holds for any value of \ibf. The corresponding value of $\ibf\deltav$, and therefore the \deltav\ we use in scenario (d), is computed using Eqs.~(\ref{eqn:vperi}) and~(\ref{eqn:vafter}).

\begin{figure}[t]
\centering
\includegraphics[width=0.49\textwidth, angle=0]{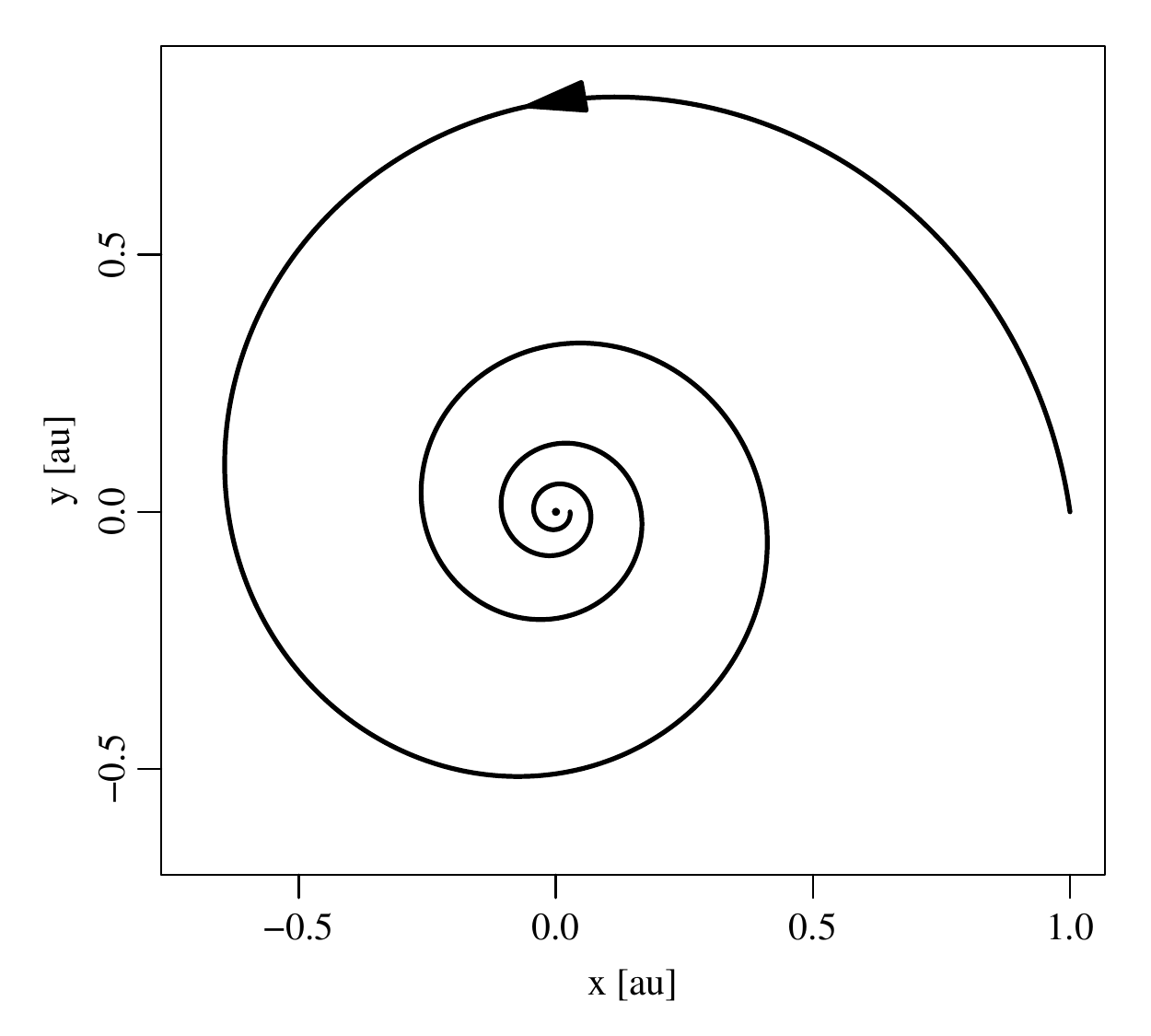}
\caption{A logarithmic spiral orbit as described by Eq.~(\ref{eqn:rspiral}) (where $x=r\cos\theta$, $y=r\sin\theta$) with $\rinit=1$\,au, $\gamma=-8.1\degrees$, shown from $\theta=0$\,rad to $\theta=8\pi$\,rad.
The dot in the center is the Sun (not plotted to scale).}
\label{fig:spiral_orbit}
\end{figure}

\begin{figure}[t]
\centering
\includegraphics[width=0.49\textwidth, angle=0]{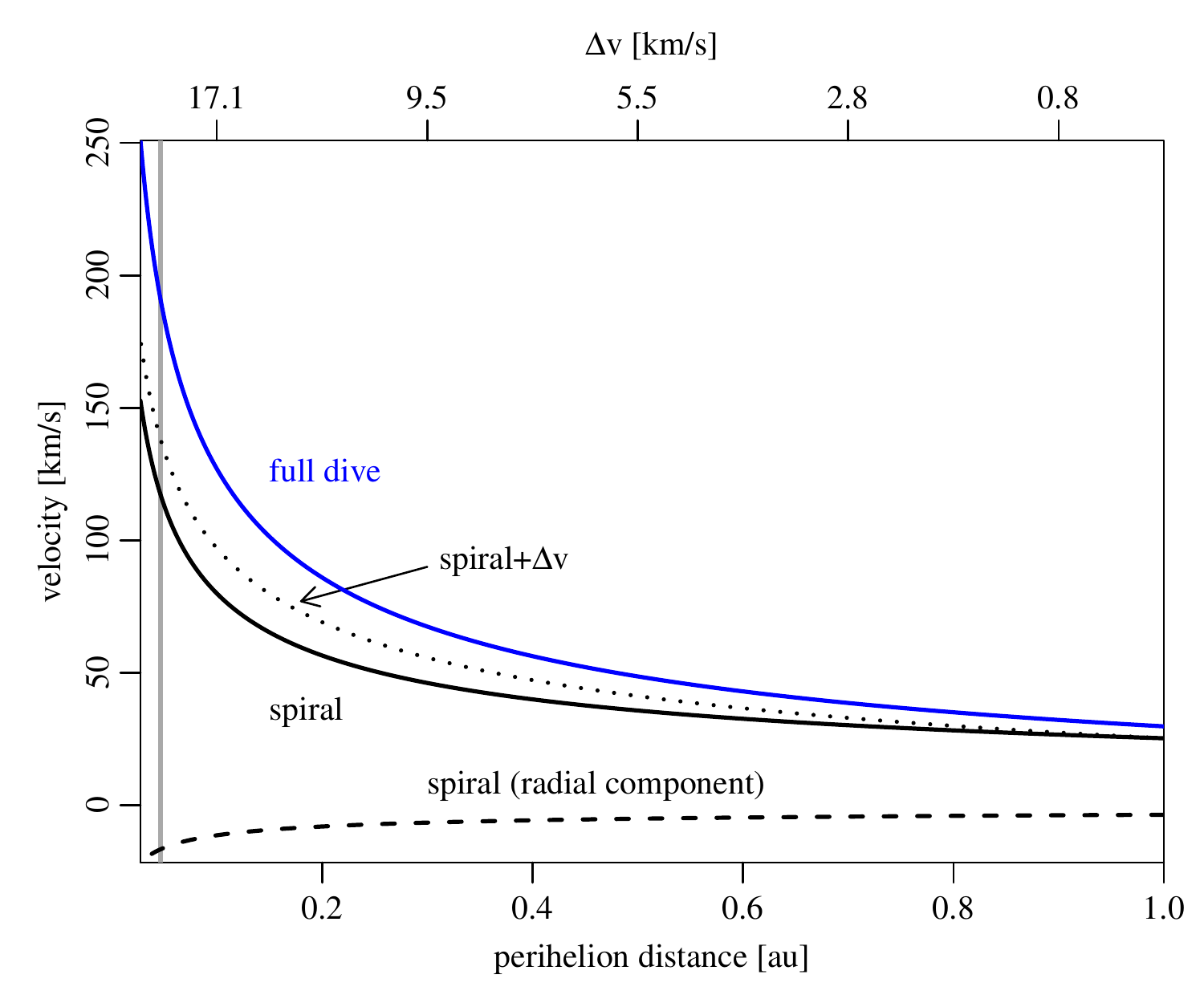}
\caption{Sail velocities as a function of perihelion distance.
The vertical solid line indicates 10 solar radii.
The lower solid line is the velocity \vspiral\ from Eq.~(\ref{eqn:vspiral}) with $r=\rperi$ for the logarithmic spiral orbit shown in Fig.~\ref{fig:spiral_orbit}. The dashed line is its radial component (which is negative). The tangential component is not show as it is almost equal to \vspiral. The dotted line is
$\vspiral + \deltav$, i.e.\ the velocity of the sail after the prograde boost (scenario (d)). The upper solid line is the perihelion velocity for the full dive (scenario (a), Eq.~(\ref{eqn:vperi_alt})), which establishes a one-to-one relationship between \rperi\ and \deltav\ (Fig.~\ref{fig:rperi_and_vperi2_vs_deltaV}) and is shown along the top axis.
We see that the velocity achieved in scenario (a) is always larger at a given \deltav\ than using this same \deltav\ in scenario (d).
}
\label{fig:spiral_and_fulldive_velocities}
\end{figure}

For the sake of illustration let us adopt the following parameters for the spiral orbit: $\lightness=0.3$ and $\alpha=-10\degrees$, which correspond to $\gamma=-8.1\degrees$. The orbit for four revolutions around the Sun is shown in Fig.~\ref{fig:spiral_orbit}.
The magnitude of the velocity, \vspiral, as a function of
radial distance is shown by the lower solid line in Fig.~\ref{fig:spiral_and_fulldive_velocities}.
For the spiral orbit of scenario (d), the closest approach velocity that we achieve is $\vspiral+\deltav$, shown by the dotted line.
For the full dive of scenario (a), the closest approach velocity is shown by the upper solid line.
We see that the closest approach velocity for scenario (a) is above that for scenario (d) for any given $r_p$ (which is equivalent to any given $\deltav$).
Although this is shown here for scenario (d) with a certain \lightness\ and $\alpha$, we find that it holds for any \lightness\ and $\alpha$.

If only a limited \deltav\ were available, say 5\,\kms, then in scenario (a) we achieve $\rperi=0.53$\,au and $\vperi=46.8$\,\kms, whereas in scenario (d) we could perhaps spiral in much closer to the Sun, e.g.\ to 10 solar radii where $\vspiral=117.2$\,\kms, and then apply the same \deltav\ to get a closest approach velocity of $122.2$\,\kms.
But the main point of the comparison in this section is to show that if we have enough \deltav\ to dive as close to the Sun as is thermally possible, then scenario (d) is always inferior to scenario (a): Spiraling towards the Sun to this distance and then applying that \deltav\ always results in a smaller velocity than a full dive.

It was mentioned above that an additional impulse would in practice be needed to put the spacecraft on a logarithmic spiral trajectory in the first place. This would take away some of the available \deltav, making scenario (d) even less favorable.



\subsection{Multi-stage transfer orbits}\label{sec:more_complex}

In section~\ref{sec:nominal_scenario} we only considered a single elliptical transfer orbit, from the initial circular orbit to the point where the sail is deployed.
%
Multi-stage transfers can also be considered, and these are sometimes used in practice because they sometimes need a smaller total \deltav\ to move between the same two orbits.  An example is the bi-elliptic Hohmann transfer orbit.\cite{curtis2014}
Starting on a circular orbit of 1\,au radius, a prograde boost is used to put the spacecraft on an elliptical orbit of higher energy and thus larger semi-major axis than the initial orbit. When the spacecraft reaches aphelion, a relatively small retrograde boost is sufficient to put the spaceraft on a less eccentric, lower energy orbit, meaning it will dive close to the Sun. A final retrograde boost could be applied at perihelion to put the spacecraft on a low circular orbit around the Sun, but in our application we would now open the solar sail to move away from the Sun at high velocity. Such a maneuver can
be set up to require less \deltav\ than our simple one-transfer maneuver to reach a given perihelion (or to reach a smaller perihelion for a given \deltav).

\section{Conclusions}

We have examined the consequences of combining impulsive boosts with solar sails in Keplerian orbits as a way of maximizing the velocity of a spacecraft at infinity.
One of the main conclusions of this study may appear counter-intuitive: decelerating a solar sail by some \deltav\ can result in a larger velocity at infinity than accelerating it by the same \deltav.
This is always the case for sufficiently large \deltav\ or sail lightness number \lightness\ (Fig.~\ref{fig:transition_lightness_vs_deltav}), and is true for any \deltav\ when $\lightness>1/2$. In these cases the largest velocity at infinity is achieved by using the entire \deltav\ in a retrograde burn to dive as close to the Sun as possible before opening the sail at perihelion.
This is because the extra energy acquired from the solar radiation by diving close to the Sun more than compensates for the energy lost by performing the dive.
For smaller \deltav\ or lightness number, a larger velocity at infinity is achieved by instead using the entire \deltav\ in a prograde burn at the moment the sail is deployed to move away from the Sun without performing any dive.
A combination of retrograde and prograde burns is always suboptimal. Tilting the sail to spiral in to the Sun before applying the \deltav, or using a sail (with lightness number above $1/2$) to travel directly to infinity and then applying \deltav, are also inferior in terms of final velocity achieved.
We have looked here only at the use of a single transfer orbit.
A natural extension of this work would be to examine the optimal combination of three impulsive boosts for two transfer orbits that maximize the spacecraft velocity at infinity.

\begin{acknowledgments}

I thank Thomas M\"uller and Markus P\"ossel for useful comments on the manuscript.

\end{acknowledgments}


\begin{thebibliography}{99}

\bibitem{walter2018} U.\ Walter, {\em Astronautics} (Springer, Cham, 2018).

\bibitem{pinheiro2004} M.J.\ Pinheiro, ``Some remarks about variable mass systems,''
Eur.\ J.\ Phys.\ \textbf{25}, L5--L7 (2004).

\bibitem{mcinnes1999} C.R.\ McInnes, {\em Solar sailing} (Springer, Heidelberg, 1999).

\bibitem{vulpetti2015} G.\ Vulpetti, L.\ Johnson, G.L.\ Matloff, {\em Solar sail: a novel approach to interplanetary travel} (Springer-Praxis, Chichester, 2015)
  
\bibitem{tsuda2013} Y.\ Tsuda, O.\ Mori, R.\ Funase, et al., ``Achievement of IKAROS — Japanese deep space solar sail demonstration mission,'' Acta Astronautica \textbf{82}, 183--188 (2013).

\bibitem{liewer2000} P.C.\ Liewer, R.A.\ Mewaldt, J.A.\ Ayon, et al., ``NASA's interstellar probe mission,''
  in {\em Space Technology and Applications International Forum-2000},  edited by M.S.\ El-Genk (AIP, 2000), p.~911.
 
\bibitem{lyngvi2005} A.\ Lyngvi, P.\ Falkner, A.\ Peacock, ``The interstellar heliopause probe technology reference study,'' Advances in Space Research \textbf{35}, 2073--2077 (2005).
  
\bibitem{gong2011} S.-P.\ Gong, Y.-F.\ Gao, J.-F.\ Li,``Solar sail time-optimal interplanetary transfer trajectory design,'' Research in Astron.\ Astrophys.\ \textbf{11}, 981--996 (2011).

\bibitem{mcnutt2003} R.L.\ McNutt Jr., G.B.\ Andrews, J.\ McAdams, et al., ``Low-cost interstellar probe,''
  Acta Astronautica \textbf{52}, 267--279 (2003).
  
\bibitem{curtis2014} H.D.\ Curtis, {\em Orbital mechanics for engineering students}, (Butterworth-Heinemann, Oxford, 2014).

\bibitem{deltav_magnitude} We examine values of \deltav\ up to nearly the Earth's orbital velocity of $\vinit=29.8$\,\kms.  The \deltav\ used by spacecraft rockets in interplanetary space are typically less than 1\,\kms, and even to launch from the Earth's surface to low Earth orbit the required \deltav\ is ``only'' around 10\,\kms.  Due to the rocket equation, achieving a large \deltav\ for our sail with chemical rockets demands very large mass ratios:  Even when using a liquid fuel rocket with one of the highest effective exhaust velocities available, 4.4\,\kms, then to attain $\deltav = 20$\,\kms\ would require a propellant to spacecraft mass ratio of about 100 (and 260 for $\deltav = 25$\,\kms).
Thus the upper end of the \deltav\ values examined should be considered theoretical rather than practically achievable today.
  
  \bibitem{betts2019} Betts B., Spencer D.A., Bellardo J.M., et al., ``Lightsail 2: Controlled solar sail propulsion using a cubesat,'' 70th International Astronautical Congress (IAC), Washington D.C., USA, October 2019.
  
\bibitem{periapsis_definition} The periapsis is the closest point on an orbit to
to the central body. When that body is the Sun, this point is called the perihelion. The furthest point on the orbit is called the apoapsis and aphelion.
  
\bibitem{oberth1929} H.\ Oberth, {\em Wege zur Raumschiffahrt} (R.\ Oldenbourg, M\"unchen--Berlin, 1929).

  \bibitem{longcope2000} D.\ Longcope, ``Using Kepler's laws and Rutherford scattering to chart the seven gravity assists in the epic sunward journey of the Parker Solar Probe,''  Am.\ J.\ Phys.\ \textbf{88}, 11--19 (2000).

\bibitem{psp}http://parkersolarprobe.jhuapl.edu

\bibitem{bacon1959} R.H.\ Bacon, ``Logarithmic spiral: An ideal trajectory for the interplanetary vehicle with engines of low sustained thrust,''  Am.\ J.\ Phys.\ \textbf{27}, 164--165 (1959).


\end{thebibliography}
\end{document}